\documentclass[aps,prl,twocolumn,groupedaddress,showpacs,superscriptaddress]{revtex4}

\usepackage{graphicx}

\begin{document}

\title{Intrinsic noise properties of atomic point contact displacement detectors}

\author{N. E. Flowers-Jacobs}
\email{nathan.flowers-jacobs@colorado.edu} \affiliation{JILA,
National Institute of Standards and Technology and the University of
Colorado, Boulder, CO 80309, USA} \affiliation{Department of
Physics, University of Colorado, Boulder, CO 80309, USA}
\author{ D. R. Schmidt}
\affiliation{JILA, National Institute of Standards and Technology
and the University of Colorado, Boulder, CO 80309, USA}
\affiliation{Department of Physics, University of Colorado, Boulder,
CO 80309, USA}
\author{K. W. Lehnert}
\affiliation{JILA, National Institute of Standards and Technology
and the University of Colorado, Boulder, CO 80309, USA}
\affiliation{Department of Physics, University of Colorado, Boulder,
CO 80309, USA}

\begin{abstract}
We measure the noise added by an atomic point contact operated as a
displacement detector. With a microwave technique, we increase the
measurement speed of atomic point contacts by a factor of 500. The
measurement is then fast enough to detect the resonant motion of a
nanomechanical beam at frequencies up to 60~MHz and sensitive enough
to observe the random thermal motion of the beam at 250~mK. We
demonstrate a shot-noise limited imprecision of $2.3\pm0.1\
\mathrm{fm/\sqrt{Hz}}$ and observe a $78\pm 20 \
\mathrm{aN/\sqrt{Hz}}$ backaction force, yielding a total
uncertainty in the beam's displacement that is $42\pm9$ times the
standard quantum limit.
\end{abstract}


\maketitle

Both to realize quantum-limited measurement of position and to
develop ultrasensitive detectors, there has been a renewed effort to
more precisely measure the motion of nanomechanical elements. Recent
examples include the demonstration of near quantum limited
displacement sensitivity using single electron transistors
\cite{knobel2003,lahaye2004} as well as novel sensors of
nanomechanical motion that use mesoscopic strain gauges in the form
of carbon nanotubes\cite{Savonova2004} or quantum point contacts
\cite{cleland2002}.  There has been a parallel development in
theories that calculate the noise properties of these sensors
\cite{mozyrsky2002, mozyrsky2004, bocko1988, yurke1990,
blencowe2000, clerk2004PRB245306}, focusing on how closely they can
approach displacement measurement at the standard quantum limit
\cite{clerk2004PRB245306,caves1982}. Reaching the standard quantum
limit is a compromise between minimizing the purely apparent motion
of the nanomechanical element inferred from noise at the output of
the sensor, that is, the displacement imprecision $S_{x}(\omega)$,
and minimizing the real random motion caused by the sensor,
described as a backaction force with spectral density
$S_{F}(\omega)$. A continuous linear measurement of position is
subject to the Heisenberg constraint $\sqrt{S_{x}S_{F}} \geq \hbar$,
Ref. \cite{spectraldensitynote}. When measuring the position of a
harmonic oscillator, the standard quantum limit is most closely
approached when $S_{x}(\omega)=|H(\omega)|^{2} S_{F}(\omega)$
assuming $S_x(\omega)$ and $S_F(\omega)$ are uncorrelated, where
$H(\omega)$ is the harmonic oscillator's response function
\cite{caves1982, clerk2004PRB245306}.

A displacement detector based on electrons tunneling from an atomic
point contact (APC) has three attributes that make it a promising
amplifier of nanomechanical motion. First, the intrinsic noise at
the output of the atomic point contact, electrical shot-noise, can
easily overwhelm the noise at the input of conventional amplifiers.
This is a crucial prerequisite for quantum-limited amplification.
Second, the APC need not be operated at ultralow temperatures. Both
advantages arise from the relatively large energy scale, the few
electron-volt work function of the metal, that controls electron
tunneling in the APC. Finally, recent theoretical work
\cite{bocko1988,yurke1990,mozyrsky2002,clerk2004PRB245306,clerk2004PRB121303}
has predicted that a displacement detector based on an APC would be
a quantum mechanically ideal amplifier. In addition, the atomic
point contact is already an important and commonly used displacement
detector as it provides the exceptional spatial resolution in
scanning tunneling microscopes and mechanically adjustable
break-junctions. Until now, the atomic point contact neither has
been operated with sufficient temporal resolution to sense a
nanomechanical element moving at its resonance frequency nor has its
backaction been measured.

In this Letter, we measure the intrinsic noise of a displacement
sensor based on an atomic point contact, finding both its
imprecision $S_x$ and backaction $S_F$ spectral densities. As in
scanning tunneling microscopy, we infer the distance between an
atomically sharp point and a nearby conducting object from the
tunneling resistance across the APC. However, by detecting the APC's
resistance at microwave frequencies we can measure displacement on
times less than 10 nanoseconds. The measurement is fast and
sensitive enough to resolve the submicrosecond, subpicometer
resonant motion of a nanomechanical beam driven by thermal noise at
temperatures below 1~K. The random thermal motion of the
nanomechanical beam provides a calibrated noise source which can be
used to determine the displacement imprecision and backaction of the
APC.  Our current realization of this displacement detector, an APC
amplifier, has a displacement imprecision $\sqrt{S_{x}}=2.3\pm0.1\
\mathrm{fm/\sqrt{Hz}}$ and a backaction $\sqrt{S_{F}}=78\pm20\
\mathrm{aN/\sqrt{Hz}}$ yielding a total displacement uncertainty
that is 42 times the standard quantum limit and an
imprecision-backaction product $\sqrt{S_{x}S_{F}}= 1700 \pm400\
\hbar$. We demonstrate that the imprecision of our measurement is
shot-noise limited; however, we also observe that the backaction
force is much larger than theoretically predicted.

\begin{figure}
\includegraphics{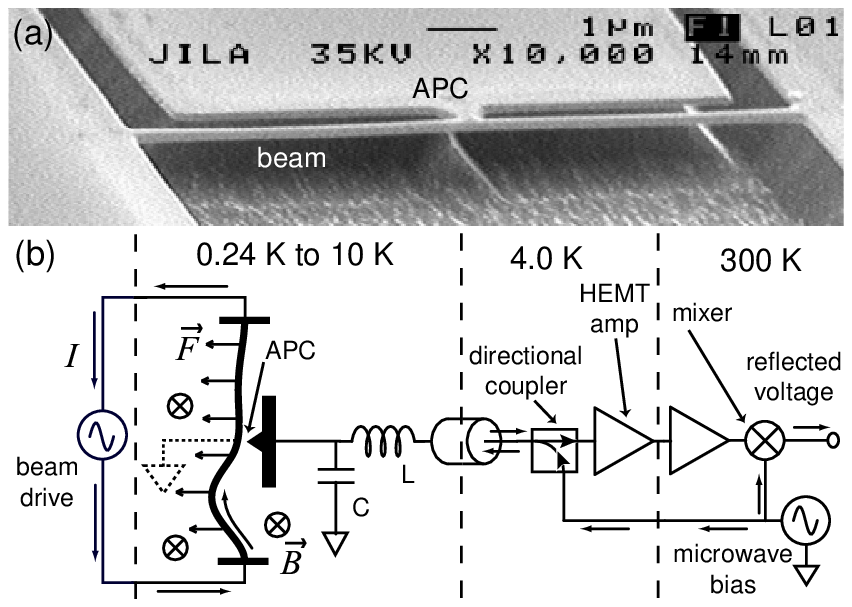}\\
\includegraphics{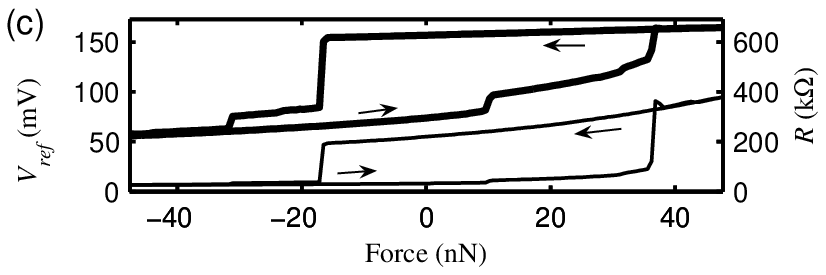}\\
\caption{\label{fig1} (a) Representative scanning electron
micrograph of the nanomechanical system consisting of a doubly
clamped beam suspended above a GaAs substrate and a triangular
electrode fused to the beam center; the APC is formed at the
junction between the electrode and the beam. (b) Simplified
schematic consisting of a displacement measurement shown to the
right of the APC and a drive mechanism (using a Lorentz force
created by passing a current through the beam in the presence of a
9~T magnetic field) shown to the left. (c) $V_{ref}$ (thick line)
and APC resistance (thin line, measured at 7 Hz using a lock-in
amplifier) versus static applied Lorentz force. We observe a
monotonic relationship between $R$ and $V_{ref}$.}
\end{figure}

An APC is formed by bringing an atomically sharp conducting point
within 1 nm of another conducting object \cite{agrait2003}.  To
monitor the size of the gap between the point and the other object
[Fig. \ref{fig1}(a)], a voltage bias is applied across the gap and
the probability of electron tunneling is determined by measuring the
current.  The ratio of the voltage $V$ to the average current
$\langle I \rangle$ is a resistance $R$ that changes monotonically
with the size of the gap. The APC can then be regarded as both a
transducer and an amplifier of small displacements $d x$ with an
output $d \langle I \rangle =G\ dx$, where $G=(1/R)(\partial R
/\partial x)\langle I \rangle$ is the amplifier's gain.

Only a small fraction of an APC's intrinsically large bandwidth is
typically realized because the APC is also necessarily a
high-resistance device \cite{agrait2003}. The high resistance
shunted by the large cable capacitance between the device and the
remote measurement electronics \cite{schoelkopf1998} creates an
approximately 100~kHz bandwidth limit. We overcome this bandwidth
limitation by transforming down the resistance of the APC towards
$50\ \Omega$ with an electrical resonant transformer formed out of
an inductor $L$ and capacitor $C$, inferring the resistance by
measuring the on-resonance ($1/2\pi \sqrt{LC} = 430\ \mathrm{MHz}$)
reflected microwave voltage $V_{ref}$ [Fig. \ref{fig1}(b,c)], where
$\langle I \rangle$ is now the magnitude of the \textit{microwave}
current passing through the APC. With this technique, originally
implemented in the radio-frequency single-electron
transistor\cite{schoelkopf1998}, we achieve a bandwidth of 30~MHz
controlled by the quality factor of the resonant circuit; it is
possible to detect motion outside of this band at the cost of a
larger contribution to the measurement noise by the conventional
electronics. Further increases in the bandwidth are possible by
using higher-frequency resonant circuits. We choose to operate the
APC amplifier with a 430~MHz resonance frequency because low-noise
microwave amplifiers are readily available at this frequency, and
the resonant circuit can be fabricated from discrete components.

The mechanical system in this experiment is composed of a doubly
clamped nanomechanical beam next to an atomically sharp point [Fig.
\ref{fig1}(a)]. The beam and point electrode are made entirely out
of gold and are fabricated fused together. An APC is formed between
the point and the beam by creating a gap using
electromigration\cite{park1999} in the ultrahigh vacuum present in a
4~K cryostat; despite this precaution, it is still possible that
contaminants in the gap between the beam and the point play a role
in this experiment. The beam is 5.6-$\mu$m long by 220-nm wide by
100-nm thick resulting in a total beam mass $m=2.3\times10^{-15}\
\mathrm{kg}$.

\begin{figure}
\includegraphics{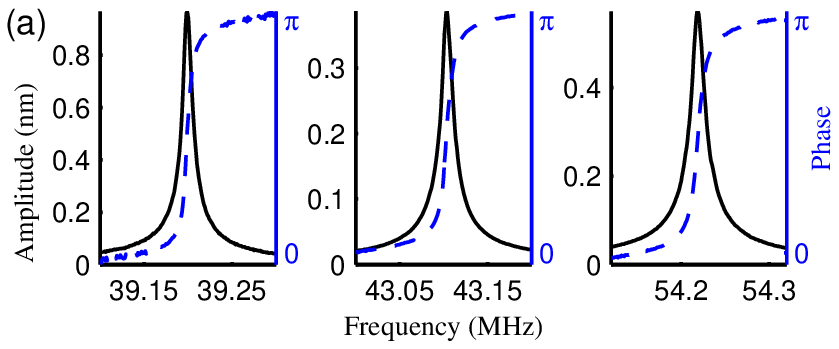}\\
\includegraphics{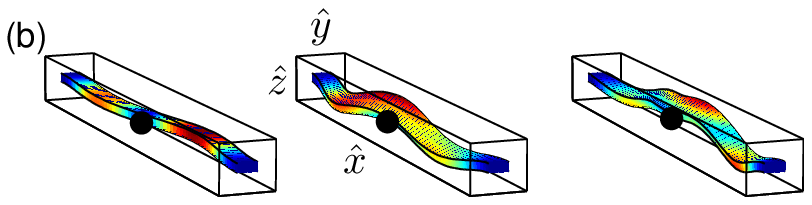}
\caption{\label{fig2}(Color online) (a) Measured response amplitude
(solid) and phase (dashed) as a function of drive frequency for
three of the five observed resonant modes measured with the APC
amplifier.  (b) Corresponding finite-element simulation of beam mode
shapes (dot corresponds to position of spring used to model the
interatomic potential at the APC; color indicates displacement from
equilibrium position with minimum displacement at the ends of the
beam).}
\end{figure}

We demonstrate the bandwidth of the APC amplifier by finding the
resonant frequencies of the beam.  These resonant modes are detected
by sweeping the frequency of an oscillating 50~pN Lorentz force
[Fig. \ref{fig1}(b)] applied to the beam parallel to the substrate
($\hat{y}$ direction, [Fig. \ref{fig2}(b)]) and using the APC
amplifier to look for resonances\cite{ekinci2005}. We observe five
resonance frequencies between $\omega_{0}/2\pi=$18.4~MHz and
57.2~MHz with a typical quality factor of 5000 [Fig. \ref{fig2}(a)].
An interatomic potential between the gold atoms comprising the APC
modifies the resonance frequencies and mode shapes from those
expected for a doubly clamped beam. While for large static applied
forces there is hysteresis in the beam's displacement [Fig.
\ref{fig1}(c)], for small displacements from mechanical equilibrium
the net effect of the interatomic potential deflecting the beam can
be modeled as a spring spanning the APC, which connects the point
and the beam. In a finite element simulation, we adjust the
compliance of the spring until the simulated resonance frequencies
match those observed. Although this 180~N/m spring results in
simulated mode shapes that appear complicated [Fig. \ref{fig2}(b)],
near a resonance frequency the system behaves like a one-dimensional
simple harmonic oscillator \cite{ekinci2005} with a collective
coordinate whose displacement is equal to the root-mean-squared
displacement of the beam averaged over the beam's length [Fig.
\ref{fig2}(a)]. Because we specify displacement exclusively in terms
of this normal (or modal) coordinate, the effective spring constant
of the simple harmonic oscillator is $k_{s}=m\omega_{0}^{2}$ with
$m$ equal to the \emph{total} mass of the beam.

\begin{figure}
\includegraphics{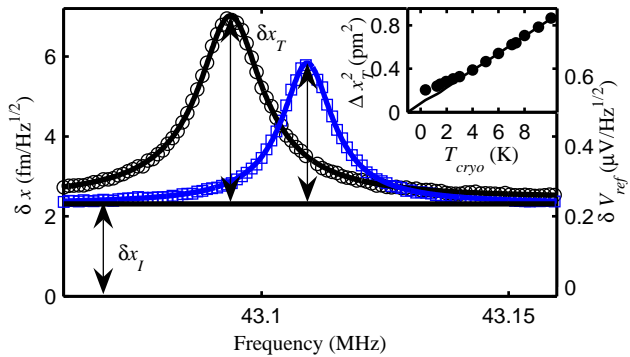}
\caption{\label{fig3}(Color online) (Main) APC amplifier noise
spectrum $\delta V_{ref}$ (right axis) versus frequency and
corresponding displacement fluctuations of the beam $\delta x$ (left
axis) at 5~K (squares, right peak) and 10~K (circles, left peak) and
square root of Lorentzian fits (lines); the noise spectra were
measured with an APC resistance of 33~k$\Omega$, and for this
$G=290$~nA/nm the microwave voltage across the APC was 26~mV.
(Inset) Integrated strength $\Delta x_{T}^2$ as a function of
$T_{cryo}$ (dot, $G=290$~nA/nm) with a fit (line) to the model for
$\Delta x_{T}^2$. For low temperatures we observe deviations from
the model due to locally heating by the bias current of the
dissipative environment coupled to the mechanical system.}
\end{figure}

\begin{figure}
\includegraphics{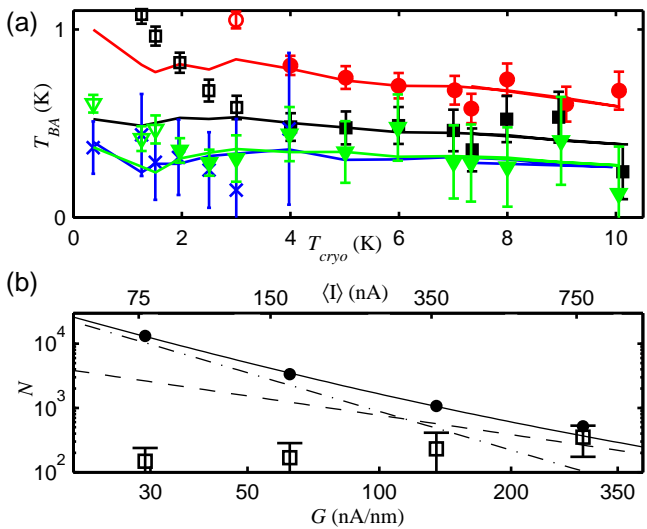}
\caption{\label{fig4}(Color online) (a) Backaction temperature
$T_{BA}$ versus the temperature of the cryogenic system, $T_{cryo}$,
with gains $G$=290~nA/nm (circle), $G$=130~nA/nm (square),
$G$=60~nA/nm (triangle), and $G$=30~nA/nm (cross).  Error bars are
extracted from one sigma uncertainties in fits to Lorentzian noise
peaks [Fig. \ref{fig3}]; hollow symbols are excluded from the fit
(lines) to the model. (b) Measurement imprecision quanta
$N_{x}=k_{s} S_{x} \gamma /4 \hbar \omega_{0}$ (solid dots) and
momentum backaction quanta $N_{p}=S_{F}/ 4 m \gamma \hbar
\omega_{0}$ (hollow squares) at 5~K as a function of $G$. The
uncertainty in the imprecision (error bars are the size of the solid
dots) and backaction are dominated by the uncertainty in the beam
temperature.  As $G$ is increased by increasing the average current
$\langle I \rangle$, the compromise between a reduced imprecision
and increased backaction is evident. At the maximum $G$ the
imprecision is determined by the intrinsic noise $S_{I}^{SN}$ of the
APC amplifier $N_{SN}= k_{s} S_{I}^{SN} \gamma /4 G^{2}\hbar
\omega_{0}$ (dashed line). For smaller values of $G$, the
imprecision is dominated by noise $S_{I}^{A}$ added by the HEMT
microwave amplifier $N_{A}=k_{s} S_{I}^{A} \gamma /4 G^{2}\hbar
\omega_{0}$ (dot-dashed line), [Fig. \ref{fig1}(c)]. The measurement
imprecision is the sum of these two contribution
$N_{x}=N_{SN}+N_{A}$ (solid line).}
\end{figure}

To characterize the displacement imprecision and amplifier
backaction in our measurement, we find the noise spectrum $\delta
V_{ref}$ when the beam is not driven by a Lorentz force [Fig.
\ref{fig3}(main)]. The noise spectrum has two components, a
frequency independent background and Lorentzian peaks at the beam's
mechanical resonance frequencies due to the Brownian motion of the
beam. We calibrate $\delta V_{ref}$ in displacement units through
the temperature dependence of the Brownian motion. Through this
ratio of $V_{ref}$ and $x$ and through the relationship between
$V_{ref}$ and $R$ in [Fig. \ref{fig1}(c)], we find $(1/R)(\partial R
/\partial x)= 0.4 \mathrm{~nm^{-1}}$ when $R=33\mathrm{~k\Omega}$,
the operating point where we observe minimum imprecision. From a fit
to the Lorentzian peak we extract the full width of the peak at half
maximum $\gamma/2\pi$ (defined so that the harmonic oscillator feels
a dissipative force $m \gamma (\partial x/\partial t)$), center
frequency $w_{0}/2\pi$, amplitude $\delta x_{T}$, and
frequency-independent background $\delta x_{I}=\sqrt{S_x}$. These
fit values are used to calculate the integrated strength of the
Lorentzian peak $\Delta x_{T}^2=\delta x_{T}^2 \gamma/4$ as a
function of the cryostat's temperature $T_{cryo}$ [Fig.
\ref{fig3}(inset)]. While the equipartition theorem predicts $\Delta
x_{T}^{2}=k_{B} T_{cryo}/k_{s}$, we model $\Delta x_{T}^{2}=(k_{B}
T_{cryo}/k_{s})+(S_F/4 m \gamma k_{s})$, where the spectral density
of a random force $S_F$ depends upon $G$ (which is equivalent to
depending on the amplitude of the microwave bias current). From this
dependence, we infer that the measurement process imposes a random
force $S_F$ effectively heating the beam to a temperature
$T_{BA}=S_F/4 m \gamma k_B$ above $T_{cryo}$ [Fig. \ref{fig4}a].
While we model $S_F$ as independent of $T_{cryo}$, $T_{BA}$ has a
$T_{cryo}$ dependence through $\gamma$, which changes by 30\% in the
0.25~K to 10~K range of $T_{cryo}$. For the maximum gain $G=290~
\mathrm{nA/nm}$ studied and $\omega_{0}/2\pi=43.1\ \mathrm{MHz}$, we
measure both the minimum imprecision $\sqrt{S_{x}}=2.3\pm0.1\
\mathrm{fm/\sqrt{Hz}}$ and the maximum random force $\sqrt{S_{F}}=
78\pm20~\mathrm{aN/\sqrt{Hz}}$ acting on the beam; this random force
implies that at $T_{cryo}=5~ \mathrm{K}$ there is a backaction
temperature $T_{BA}=0.73\pm0.4~ \mathrm{K}$.

We compare these results to limits imposed by the Heisenberg
uncertainty principle.  A Heisenberg limited amplifier with gain
chosen to operate at the standard quantum limit would have an
imprecision $S_{x}^{SQL}= \hbar |H(\omega)|$ and a backaction force
$S_{F}^{SQL}= \hbar /|H(\omega)|$, where
$H(\omega)=1/m(\omega_0^2-\omega^2+i\omega\gamma)$ is the beam's
response function \cite{caves1982, clerk2004PRB245306, ekinci2005}.
On resonance, $\sqrt{S_{x}^{SQL}}=\sqrt{\hbar / m \omega_{0}
\gamma}=51\ \mathrm{am/\sqrt{Hz}}$ and $\sqrt{S_{F}^{SQL}}=
\sqrt{\hbar m \omega_{0} \gamma}= 2.1~\mathrm{aN/\sqrt{Hz}}$.
Comparing to these, we find that for our APC amplifier
$\sqrt{S_{x}}=45\ \sqrt{S_{x}^{SQL}}$ and
$\sqrt{S_{F}}=38~\sqrt{S_{F}^{SQL}}$. By assuming that the
fluctuations imposed by the random backaction force are uncorrelated
with the measurement imprecision, we make a worst case estimate of
the quantum nonideality of the APC as a displacement amplifier. We
find that $\sqrt{S_{x}S_{F}}=1700\pm400\ \hbar$, which is 1700 times
the Heisenberg-limited imprecision-backaction product. As with most
amplifiers, the APC amplifier imprecision can be reduced by
increasing $G$ at the expense of a larger backaction
\cite{caves1982, clerk2004PRB245306} [Fig. \ref{fig4}(b)]. We have
therefore operated the APC amplifier near an optimum $G$ that
minimizes the {\emph{total}} displacement uncertainty on resonance
$\sqrt{S_{x}^{tot}}=\sqrt{S_{x}+S_{F} |H(\omega_0)|^2} =42\pm
9~\sqrt{2 \hbar / m \omega_{0} \gamma}$, i.e., $42$ times the total
displacement uncertainty at the standard quantum limit [Fig.
\ref{fig4}].

Because the physical origins of the imprecision and backaction that
combine to enforce the Heisenberg limit in the theoretical treatment
of an APC amplifier are readily understood
\cite{bocko1988,yurke1990}, we can speculate about the source of the
nonideality in our APC amplifier. The fundamental imprecision in the
APC is electrical shot noise; that is, random Poissonian
fluctuations in the electron tunneling rate
\cite{bocko1988,clerk2004PRB121303}. Similarly, the backaction that
enforces the Heisenberg limit is due to the random magnitude of the
momentum kicks imparted to the beam by each electron that tunnels
\cite{yurke1990}. To realize an ideal quantum amplifier, these
sources of imprecision and backaction must dominate. In contrast to
single-electron transistor and quantum point contact displacement
detection \cite{knobel2003,lahaye2004,cleland2002}, it is relatively
easy to ensure that the noise power $S_{I}^{A}$ added by
conventional amplifiers and electrical components [Fig.
\ref{fig1}(b)] is overwhelmed by the shot noise of tunneling
electrons $S_{I}^{SN}=2e \langle I \rangle(2 \sqrt{2}/\pi)$, where
the $2 \sqrt{2}/\pi$ factor arises from averaging the instantaneous
shot noise over one cycle of the microwave bias. Indeed we find that
for the $G$=290~nA/nm, the maximum gain studied, the electrical shot
noise accounts for 70\% of the spectral density of the displacement
imprecision and only 30\% comes from the rest of the measurement
circuit [Fig. \ref{fig4}(b)].

In spite of the fact that the imprecision is shot-noise limited, it
is still substantially larger than was estimated by
\cite{bocko1988}. It is through a smaller value of $\partial
R/\partial x$ that the gain $G$ at a particular current value is
smaller and the shot-noise limited imprecision
$S_{x}=S_{I}^{SN}/G^2$ is larger than predicted. Assuming the APC
resistance obeys $R=R_{0} \exp(2 \alpha x /\lambda)$, the usual
one-dimensional model for electron tunneling \cite{simmons1693},
where $\lambda = 90\mathrm{~pm}$ in gold and $\alpha=0.4 \pm 0.2$ is
a mode dependent factor relating normal coordinate displacement $x$
to displacement at the APC, then $\sqrt{S_{x}}$ should be $70\
\mathrm{am/\sqrt{Hz}}$ at the highest experimental bias current.
However, in our experiment we infer from $\partial R/\partial x$
that $\lambda=2 \pm 1\ \mathrm{nm}$ resulting in a shot-noise
limited imprecision $\sqrt{S_{x}}=2\pm1\ \mathrm{fm/\sqrt{Hz}}$.
Despite known deviations from the simple exponential dependence of
resistance on position for tunnel resistances $R_{0}$ less than
1~M$\Omega$ \cite{sun2005}, we operate at resistances less than
100~k$\Omega$ so that the imprecision, displacement due to
backaction, and thermal motion (at temperatures where the bias
heating can be ignored) are all approximately equal.

Since the imprecision is shot-noise limited, we conclude that the
likely source of nonideality in our APC amplifier is a backaction in
excess of the backaction required by the Heisenberg uncertainty
principle. Two sources of excess backaction force on the
nanomechanical beam were considered by \cite{stephenson1989}. First,
shot noise in the mean momentum of the tunneling electrons results
in a backaction force. Second, an electrostatic attraction between
the beam and the APC creates a backaction force through a mutual
capacitance and a fluctuating voltage across the APC. We do not yet
understand the origin of the excess backaction; however, to account
for the observed backaction the mean momentum $\langle p \rangle$
imparted by each tunneling electron would have to be $\langle p
\rangle=e\sqrt{S_{F}/S_{I}^{SN}}=20$ times the Fermi momentum, while
the capacitive mechanism predicts a cubic dependence
\cite{bocko1988} of $S_F$ on gain $G$, apparently inconsistent with
the observed dependence [Fig. \ref{fig4}b].

In conclusion, we have demonstrated a method of using an APC to
measure displacement.  We increase the electrical bandwidth (the
measurement speed) of the APC and use this bandwidth to sensitively
detect the motion of a nanomechanical beam at frequencies up to
60~MHz. We observe the imprecision and backaction to set an upper
limit on the nonideality of an APC as a quantum amplifier, yielding
an imprecision-backaction product of $\sqrt{S_{x}S_{F}} = 1700\hbar
$ and a total displacement uncertainty of $42$ times the
standard-quantum limit. Since the imprecision of the APC amplifier
is limited by it's fundamental noise source, the shot-noise of
tunneling electrons, progress towards the quantum limit will not
require an improvement in the performance of conventional
electronics but rather an understanding of the excess backaction.

\begin{acknowledgments}
The authors acknowledge support from the National Science Foundation
under grant number PHY0096822 and the National Institute of
Standards and Technology. The authors also wish to thank J. G. E.
Harris, A. A. Clerk, S. M. Girvin, R. Mirin, and C. T. Rogers for
enlightening conversations and technical assistance.
\end{acknowledgments}


\end{document}